\documentclass[english]{IEEEtran}
\usepackage[T1]{fontenc}
\usepackage[latin9]{inputenc}
\usepackage{color}
\usepackage{float}
\usepackage{amsmath}
\usepackage{amssymb}
\usepackage{graphicx}

\makeatletter

\providecommand{\tabularnewline}{\\}
\floatstyle{ruled}
\newfloat{algorithm}{tbp}{loa}
\providecommand{\algorithmname}{Algorithm}
\floatname{algorithm}{\protect\algorithmname}

\makeatother

\usepackage{babel}
\begin{document}
\title{Reconfigurable Intelligent Surface-Assisted Uplink Sparse Code Multiple
Access}
\author{Ibrahim Al-Nahhal, \textit{Member, IEEE},\textit{ }Octavia A. Dobre,
\textit{Fellow, IEEE}, and Ertugrul Basar, \textit{Senior Member,
IEEE} \thanks{O. A. Dobre and I. Al-Nahhal are with the Faculty of Engineering and
Applied Science, Memorial University, St. John\textquoteright s, NL,
Canada, (e-mail: \{odobre, ioalnahhal\}@mun.ca).}\thanks{E. Basar is with the CoreLab, Department of Electrical and Electronics
Engineering, Ko\c{c} University, Istanbul, Turkey (e-mail: ebasar@ku.edu.tr).}\\
\vspace{-3mm}
}
\maketitle
\begin{abstract}
Reconfigurable intelligent surface-empowered communication (RIS) and
sparse code multiple access (SCMA) are promising candidates for future
generations of wireless networks. The former enhances the transmission
environments, whereas the latter provides a high spectral efficiency
transmission. This letter proposes, for the first time, an RIS-assisted
uplink SCMA (SCMA-RIS) scheme to improve the conventional SCMA spectrum
efficiency. The message passing algorithm (MPA) is utilized and modified
to decode the SCMA-RIS transmitted signals. Moreover, a low-complexity
decoder for the SCMA-RIS scheme is proposed to significantly reduce
the MPA decoding complexity and improve the bit error rate performance
of the conventional SCMA. Monte-Carlo simulations and complexity analysis
are presented, which support the findings.
\end{abstract}


\section{Introduction}

\def\figurename{Fig.}
\def\tablename{TABLE}

\IEEEPARstart{S}{parse} code multiple access (SCMA) is a code-domain
non-orthogonal multiple access (C-NOMA) approach; it represents a
promising candidate for beyond 5G wireless networks that can provide
a high spectral efficiency transmission \cite{Mohammadkarimi_Octavia}-\cite{Low-cost}.
In the SCMA scheme, a unique multi-carrier sparse code is assigned
to each user to share the wireless medium \cite{Nikopour_SCMA_2013,codebook_design_2014}.
The message passing algorithm (MPA) can be utilized to decode the
SCMA transmitted signals due to the sparsity property of the codes.
The MPA is an iterative algorithm that provides a near maximum likelihood
(ML) bit error rate (BER) performance with a relatively lower decoding
complexity \cite{MPA_2015}.

Reconfigurable intelligent surface (RIS)-empowered communication is
another promising technology for beyond 5G wireless networks, which
enhances the transmission environment for wireless communication schemes
\cite{RIS_Ertugrul 2019,RIS_2020}. Without a need for coding or encoding,
an RIS adjusts the incident signals' phases using low-cost passive
reflecting elements to enhance the signal quality at the receiver
side \cite{RIS_3}. Recently, RIS and power-domain NOMA have been
jointly explored in \cite{NOMA_RIS_1}-\cite{NOMA_RIS_3}. It is worth
noting that the power-domain and C-NOMA are completely different medium
access schemes, which share the users' data using powers and codes,
respectively. To the best of the authors' knowledge, C-NOMA, especially
SCMA, has not been investigated yet under the RIS scenario.

For the first time, this letter explores the uplink SCMA scheme assisted
by an RIS (SCMA-RIS), to improve the spectral efficiency of the conventional
SCMA scheme. The conventional ML and MPA decoders are adapted to decode
the proposed SCMA-RIS transmitted signals. Furthermore, a novel non-iterative
low-complexity (LC) decoder for the SCMA-RIS, referred to as SCMA-RIS-LC,
is proposed to overcome the high decoding complexity of the iterative
MPA. Compared to the conventional SCMA-MPA scheme, SCMA-RIS-MPA provides
significantly enhanced BER performance with slight increase in the
decoding complexity, while the SCMA-RIS-LC shows its superiority in
both BER performance and decoding complexity. Monte-Carlo simulations
and complexity analysis are provided to evaluate our findings.

The rest of the letter is organized as follows: Sections \ref{sec:Uplink-SCMA-RIS-System}
and \ref{sec:Uplink-SCMA-RIS-Signal} introduce the system model and
the detection algorithms for the proposed uplink SCMA-RIS, respectively.
In Section \ref{sec:Simulation-Results}, simulation results are presented.
Finally, Section \ref{sec:Conclusion} concludes the letter.

\section{\label{sec:Uplink-SCMA-RIS-System}Uplink SCMA-RIS System Model}

\begin{figure*}
\begin{centering}
\includegraphics[scale=0.35]{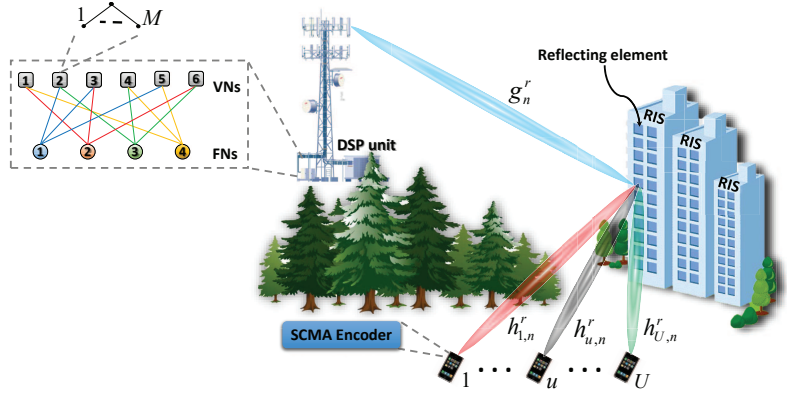}
\par\end{centering}
\vspace{-3mm}

\caption{\textcolor{blue}{\label{fig: SCMA_RIS_BlockDiagram}}Uplink SCMA-RIS
scheme.}
\end{figure*}

Consider $U$ users which deliver their data to the base station using
the SCMA scheme, through an RIS with $N$ reflecting elements, as
illustrated in Fig. \ref{fig: SCMA_RIS_BlockDiagram}. Each user accesses
the medium by employing a unique sparse codebook, $\mathbf{C}_{u}\in\mathbb{C}^{R\times M}$,
$u=1,\ldots,U$, which is spread over $R$ orthogonal resource elements
(OREs). It is worth noting that each user's codebook contains $M$
codewords, $\mathbf{c}_{u,m}\in\mathbb{C}^{R\times1}$, $m=1,\ldots,M$,
which have $d_{v}$ non-zero codeword elements. The $U$ users are
overloaded over OREs such that the number of users that share each
ORE, $d_{f}$, is fixed.

As shown in Fig. \ref{fig: SCMA_RIS_BlockDiagram}, each user transmits
a codeword, $\mathbf{c}_{u,m}$, which corresponds to $\text{lo\ensuremath{\text{g}_{2}}}(M)$
bits through $N$ passive reflecting elements. The task of the RIS
elements is to adjust the phase of the incident signal to improve
the signal quality at the receiver side. The noisy signal at the receiver
side for each ORE, $y^{r}$, is

\begin{equation}
y^{r}=\sum_{u\in\varLambda_{r}}\left(\sum_{n=1}^{N}h_{u,n}^{r}e^{\jmath\phi_{n}^{r}}g_{n}^{r}\right)c_{u,m}^{r}+w^{r},\label{eq: y_r}
\end{equation}

\noindent where $h_{u,n}^{r}$ and $g_{n}^{r}$ represent the Rayleigh
fading channel coefficient for the $r$-th ORE from the $u$-th user
to the $n$-th reflecting element and from the $n$-th reflecting
element to the receiver, respectively. Here, $\varLambda_{r}$ represents
the set of users' indices that share the $r$-th ORE, $\phi_{n}^{r}$
denotes the adjustable phase produced by the $n$-th reflecting element
at the $r$-th ORE, and $w^{r}\sim\mathcal{N}(0,\sigma^{2})$ is the
additive white Gaussian noise at the $r$-th ORE with zero-mean and
variance of $\sigma^{2}$. It is worth noting that since the OREs
for the SCMA can be orthogonal time or frequency elements, the assumption
of having an adjustable phase for each ORE is feasible.

The channel coefficients, $h_{u,n}^{r}$ and $g_{u}^{r}$, can be
written in the polar form, respectively as

\vspace{-2mm}

\begin{equation}
h_{u,n}^{r}=\alpha_{u,n}^{r}e^{-\jmath(\theta_{u,n}^{r})},\,\,\,\,\,\,\,g_{n}^{r}=\beta_{n}^{r}e^{-\jmath(\vartheta_{n}^{r})},\label{eq: h and g}
\end{equation}

\noindent Hence, by plugging (\ref{eq: h and g}) into (\ref{eq: y_r}),
the received signal, $y^{r}$, can be expressed as

\vspace{-3mm}

\begin{equation}
y^{r}=\sum_{u\in\varLambda_{r}}\left(\sum_{n=1}^{N}\alpha_{u,n}^{r}\beta_{n}^{r}e^{\jmath(\Theta_{u,n}^{r})}\right)c_{u,m}^{r}+w^{r},\label{eq: y_n_r}
\end{equation}

\noindent where $\Theta_{u,n}^{r}=\phi_{n}^{r}-(\theta_{u,n}^{r}+\vartheta_{n}^{r})$
represents the resulted phase after the adjustment of the $n$-th
reflecting RIS element for the $u$-th user at the $r$-th ORE. To
improve the instantaneous signal-to-noise ratio (SNR) at the $r$-th
ORE in (\ref{eq: y_n_r}), $\phi_{n}^{r}$ can be obtained as

\vspace{-3mm}

\begin{equation}
\phi_{n}^{r}=\underset{\begin{array}{c}
\phi_{n}^{r}\in[-\pi,\pi]\end{array}}{\text{arg}\,\text{min}}\sum_{u\in\varLambda_{r}}\left|\phi_{n}^{r}-\left(\theta_{u,n}^{r}+\vartheta_{n}^{r}\right)\right|^{2}.\label{eq: Optimiztion_min}
\end{equation}

\noindent The optimum solution for (\ref{eq: Optimiztion_min}) is
the \textit{geometric median} of the users' phases, which share that
ORE, i.e., $\phi_{n}^{r}=\text{med }(\theta_{u,n}^{r}+\vartheta_{n}^{r})$
for all $u\in\varLambda_{r}$, with $\text{med}(\centerdot)$ denoting
the geometric median. Consequently, the values of $\theta_{u,n}^{r}$
and $\vartheta_{n}^{r}$ should be known at RIS to optimize $\phi_{n}^{r}$.
To notice the improvement in the BER performance after optimizing
$\phi_{n}^{r}$, we consider a blind scenario in which the reflector
elements of RIS are reflecting the incident signals without any adjustment
to the phases (i.e., $\phi_{n}^{r}=0$ for $\forall n$). The two
scenarios of the $\phi_{n}^{r}$ can be summarized as

\vspace{1mm}

\noindent 
\begin{equation}
\phi_{n}^{r}=\begin{cases}
\begin{array}{c}
\hspace{-1.7mm}\text{med}(\theta_{u,n}^{r}+\vartheta_{n}^{r})\,\,\,\,\forall u\in\varLambda_{r}\\
0
\end{array} & \begin{array}{c}
\text{optimized}\\
\text{blind}.
\end{array}\end{cases}\label{eq: theta}
\end{equation}

The vector of received signals, $\mathbf{y}$, for the SCMA-RIS scheme
for all OREs is

\vspace{-3mm}

\begin{equation}
\mathbf{y}=\sum_{u=1}^{U}\left(\mathrm{diag}\left(\mathcal{\boldsymbol{H}}_{u}\right)\mathbf{c}_{u,m}\right)+\mathbf{w},\label{eq: y}
\end{equation}

\noindent where $\mathbf{y}\in\mathbb{C}^{R\times1}=\left[y^{1},\ldots,y^{R}\right]^{\text{T}}$,
$\mathbf{w}\in\mathbb{C}^{R\times1}=\left[w^{1},\ldots,w^{R}\right]^{\text{T}}$,
and $\mathcal{\boldsymbol{H}}_{u}\in\mathbb{C}^{R\times1}$ is

\vspace{-4mm}

\begin{equation}
\mathcal{\boldsymbol{H}}_{u}\hspace{-1mm}=\hspace{-1mm}\left[\hspace{-1mm}\left(\sum_{n=1}^{N}\alpha_{u,n}^{1}\beta_{n}^{1}e^{\jmath(\Theta_{u,n}^{1})}\hspace{-1mm}\right),\ldots,\left(\sum_{n=1}^{N}\alpha_{u,n}^{R}\beta_{n}^{R}e^{\jmath(\Theta_{u,n}^{R})}\right)\hspace{-1mm}\right]^{\text{T}}\hspace{-1mm}\hspace{-1mm},
\end{equation}

\noindent where $\mathrm{diag}(\mathcal{\boldsymbol{H}}_{u})\in\mathbb{C}^{R\times R}$
represents a diagonal matrix whose $r$-th diagonal element is $\sum_{n=1}^{N}\alpha_{u,n}^{r}\beta_{n}^{r}e^{\jmath(\Theta_{u,n}^{r})}$.

\section{\label{sec:Uplink-SCMA-RIS-Signal}Uplink SCMA-RIS Signal Detection}

In this section, we presents three decoders for the proposed uplink
SCMA-RIS scheme, which are ML, MPA and LC decoders. The complexity
analysis is also provided.

\subsection{SCMA-RIS-ML Decoder}

In SCMA-RIS, the ML decoder provides theoretical the optimum BER performance
by performing an exhaustive search for all $\left(M\right)^{U}$ users'
codeword combinations. The estimated transmitted codewords for all
users, $\hat{\mathbf{C}}_{\text{ML}}\in\mathbb{C}^{R\times U}=[\hat{\mathbf{c}}_{1,m}\ldots\hat{\mathbf{c}}_{U,m}]$,
using the ML decoder can be expressed as

\vspace{-3mm}

\begin{equation}
\hat{\mathbf{C}}_{\text{ML}}=\underset{\begin{array}{c}
j\in M^{U}\end{array}}{\text{arg}\,\text{min}}\left\Vert \mathbf{y}-\sum_{u=1}^{U}\left(\text{diag}\left(\mathcal{\boldsymbol{H}}_{u}\right)\mathbf{c}_{u,m(j)}\right)\right\Vert ^{2},\label{eq:ML-1}
\end{equation}

\noindent where $m(j)$ is the value of $m\in\left\{ 1,\ldots,M\right\} $
at the $\left(M\right)^{U}$ codeword combination. However, the SCMA-RIS-ML
is not used due to its impractical implementation.

\subsection{SCMA-RIS-MPA Decoder}

The MPA is an iterative decoder that provides a near ML BER performance
at an implementable decoding complexity for the SCMA-RIS scheme. The
SCMA-RIS-MPA uses a factor graph method as depicted in Fig. \ref{fig: SCMA_RIS_BlockDiagram}.
The MPA iteratively updates the probability of the messages between
the function nodes (FNs) representing the OREs and variable nodes
(VNs) that represent the served users. The MPA stops after $K$ iterations
to detect the users' codewords that correspond to the maximum joint
message probability. It is worth noting that the conventional SCMA-MPA
is modified to estimate the SCMA-RIS transmitted codewords.

To formulate the MPA, consider that $\mathcal{P}_{v_{u}\rightarrow f_{r}}^{(t)}(c_{m}^{r,u})$
and $\mathcal{P}_{f_{r}\rightarrow v_{u}}^{(t)}(c_{m}^{r,u})$ represent
the probability of passing the message from the $u$-th VN to the
$r$-th FN and from the $r$-th FN to the $u$-th VN, respectively,
at the $k$-th iteration, $k=1,\ldots,K$. At the first iteration,
all messages sent from VNs to FNs are assumed equiprobable; i.e.,

\vspace{-2mm}

\begin{equation}
\mathcal{P}_{v_{u}\rightarrow f_{r}}^{(0)}\left(c_{m}^{r,u}\right)=\frac{1}{M},\,\,\,\,\,\forall u,\,\,\forall r,\,\,\forall m.\label{eq:Equal_prob}
\end{equation}

\noindent Thus, $\mathcal{P}_{f_{r}\rightarrow v_{u}}^{(k+1)}(c_{m}^{r,u})$
can be written as

\vspace{-3mm}

\[
\mathcal{P}_{f_{r}\rightarrow v_{u}}^{(k+1)}\left(c_{m}^{r,u}\right)=\sum_{\psi(i),i\in\left.\varLambda_{r}\right\backslash u}\Biggl\{\mathcal{P}\left(\mathbf{y}|\psi(i),\psi(u)=c_{m}^{r,u}\right)
\]

\begin{equation}
\times\prod_{i\in\left.\varLambda_{r}\right\backslash u}\mathcal{P}_{v_{i}\rightarrow f_{r}}^{(k)}\left(\psi(i)\right)\Biggr\},\,\,\,\,\,\forall m,\,\,\forall r,\,\,u\in\varLambda_{r},\label{eq:FN_to_VN}
\end{equation}

\noindent where $\left.\varLambda_{r}\right\backslash u$ represents
$\varLambda_{r}$ except the $u$-th user, $\boldsymbol{\psi}^{r}=\{\psi(1),\ldots,\psi(i),\,\ldots\}$
denotes the possible codewords of all users that share the $r$-th
ORE, and

\vspace{-3mm}

\[
\mathcal{P}\left(\mathbf{y}|\boldsymbol{\psi}^{r}\right)=\frac{1}{\sqrt{2\pi}\sigma}\hspace{6cm}
\]

\vspace{-3mm}

\begin{equation}
\times\text{exp}\Biggl(\hspace{-0.1cm}-\frac{\left|y^{r}-\hspace{-0.1cm}\sum_{u\in\varLambda_{r}}\hspace{-0.1cm}\left(\left(\sum_{n=1}^{N}\alpha_{u,n}^{r}\beta_{n}^{r}e^{\jmath(\Theta_{u,n}^{r})}\right)c_{m}^{r,u}\right)\right|^{2}}{2\sigma^{2}}\Biggr).\label{eq:Condtional_Prob}
\end{equation}

\noindent It should be noted that the effect of RIS obviously appears
in (\ref{eq:Condtional_Prob}). Now, $\mathcal{P}_{v_{u}\rightarrow f_{r}}^{(k+1)}(c_{m}^{r,u})$
can be updated as

\[
\mathcal{P}_{v_{u}\rightarrow f_{r}}^{(k+1)}\left(c_{m}^{r,u}\right)=\gamma_{u,r}^{(k+1)}\hspace{6cm}
\]

\begin{equation}
\times\prod_{j\in\left.\Omega_{u}\right\backslash r}\hspace{-0.2cm}\mathcal{P}_{f_{r}\rightarrow v_{u}}^{(k+1)}\left(c_{m}^{r,u}\right),\hspace{0.5cm}\forall m,\,\,\forall u,\,\,r\in\Omega_{u},\label{eq:VN_to_FN-1}
\end{equation}

\noindent where $\Omega_{u}$ represents the OREs' indices that correspond
to $d_{v}$ non-zero positions of the $u$-th user, $\left.\Omega_{u}\right\backslash r$
denotes $\Omega_{u}$ except the $r$-th ORE, and $\gamma_{u,r}^{(k+1)}$
is

\vspace{-3mm}

\begin{equation}
\gamma_{u,r}^{(k+1)}=\left(\sum_{m=1}^{M}\mathcal{P}_{v_{u}\rightarrow f_{r}}^{(k)}\left(c_{m}^{r,u}\right)\right)^{-1}.\label{eq: normalization factor}
\end{equation}

After $K$ iterations, the estimated transmitted codeword of the $u$-th
user can be given by

\begin{equation}
\left\{ \hat{\mathbf{c}}_{m}^{u}\right\} ^{(K)}=\hspace{-0.4cm}\hspace{-0.2cm}\underset{\begin{array}{c}
m=1,\ldots,M\end{array}}{\text{arg}\,\text{max}}\hspace{-0.2cm}\prod_{j\in\Omega_{u}}\mathcal{P}_{f_{j}\rightarrow v_{u}}^{(K)}\left(c_{m}^{r,u}\right),\,\,\,\forall u.\label{eq: Final_MPA-1}
\end{equation}

\noindent The set of all estimated transmitted users' codewords using
the SCMA-RIS-MPA decoder, $\hat{\mathbf{C}}_{\text{MPA}}$, is

\begin{equation}
\hat{\mathbf{C}}_{\text{MPA}}=\left\{ \left\{ \hat{\mathbf{c}}_{m}^{1}\right\} ^{(K)},\,\ldots,\,\{\hat{\mathbf{c}}_{m}^{U},\}^{(K)}\right\} .\label{eq: Theta MPA}
\end{equation}

\subsection{SCMA-RIS-LC Decoder}

In this subsection, an LC algorithm for the SCMA-RIS scheme is proposed
and analyzed. The SCMA-RIS-LC is a non-iterative decoder that significantly
reduces the decoding complexity to overcome the high complexity of
the MPA. The proposed SCMA-RIS-LC decoder performs two stages to decode
the transmitted users' codewords, as follows:

\textbf{Stage 1}: In this stage, the SCMA-RIS-LC decoder detects the
transmitted users' codewords using  a single ORE by trying all possible
combinations between the users' codewords, which share that ORE. The
detected users' codewords are employed to sequentially detect the
rest of the users' codewords for the other OREs. This stage stops
when all users' codewords are detected. It should be noted that the
detected users' codewords from this stage represent the initial values
for the next stage. The detected users' codewords from Stage 1, $\hat{\mathbf{C}}_{\text{LC}}$,
is

\vspace{-3mm}

\begin{equation}
\hat{\mathbf{C}}_{\text{LC}}=\hspace{-2mm}\hspace{-2mm}\underset{\begin{array}{c}
j=1,\ldots,M^{\grave{U}^{r}}\end{array}}{\text{arg}\,\text{\,min}}\hspace{-2mm}\Bigl|y^{r}-\Psi_{1}-\Psi_{2}(j)\Bigr|^{2},\,\,\,\,1\leq r\leq R,\label{eq: Stage 1}
\end{equation}

\noindent where $\Psi_{1}$ and $\Psi_{2}$ respectively represent
the users' signals that have already been detected and that need to
be detected at the $r$-th ORE, given as

\vspace{-3mm}

\begin{equation}
\Psi_{1}=\hspace{-2mm}\sum_{u\in\grave{\varLambda}_{r}}\left(\sum_{n=1}^{N}\alpha_{u,n}^{r}\beta_{n}^{r}e^{\jmath(\Theta_{u,n}^{r})}\right)\hat{c}_{u,m}^{r}\,\Bigl|\,\hat{c}_{u,m}^{r}\subset\hat{\mathbf{C}}_{\text{LC}},\label{eq: epsi 1}
\end{equation}

\vspace{-3mm}

\begin{equation}
\Psi_{2}(j)=\sum_{u\in\varLambda_{r}\backslash\grave{\varLambda}_{r}}\left(\sum_{n=1}^{N}\alpha_{u,n}^{r}\beta_{n}^{r}e^{\jmath(\Theta_{u,n}^{r})}\right)c_{u,m(j)}^{r}.\label{eq: epsi 2}
\end{equation}

\noindent Here, $\grave{\varLambda}_{r}$ denotes the users' indices
that share the $r$-th ORE and their codewords are already detected,
$\varLambda_{r}\backslash\grave{\varLambda}_{r}$ represents $\varLambda_{r}$
except $\grave{\varLambda}_{r}$ (i.e., the user's indices whose codewords
need to be detected), and $\grave{U}^{r}=\text{card}\{\varLambda_{r}\backslash\grave{\varLambda}_{r}\}\leq d_{f}$
with $\text{card}\{\centerdot\}$ as the number of set elements.

\begin{table*}
\caption{\label{tab:Complexity_SCMA_RIS}{\small{}The decoding complexity of
the SCMA-RIS decoders}.}

\centering{}%
\begin{tabular}{c|c|c}
\cline{2-3} \cline{3-3} 
 & Real Additions (RA) & Real Multiplications (RM)\tabularnewline
\hline 
\hline 
SCMA-MPA & $\begin{array}{c}
Rd_{f}M^{d_{f}}\left(4d_{f}+K+1\right)-KRd_{f}\end{array}$ & $\begin{array}{c}
Rd_{f}M^{d_{f}}\left(4d_{f}+Kd_{f}+3\right)+M\left(d_{v}-1\right)\left(KRd_{f}+U\right)\end{array}$\tabularnewline
\hline 
$\hspace{-1mm}\hspace{-1mm}$SCMA-RIS-MPA$\hspace{-1mm}\hspace{-1mm}$ & $\begin{array}{c}
Rd_{f}M^{d_{f}}\left(4d_{f}+K+1\right)+Rd_{f}(N-K)\end{array}+1$ & $\begin{array}{c}
\hspace{-3.5mm}Rd_{f}M^{d_{f}}\hspace{-1mm}\left(4d_{f}+Kd_{f}+3\right)\hspace{-0.5mm}+\hspace{-0.5mm}M\left(d_{v}-1\right)\left(KRd_{f}+U\right)\end{array}\hspace{-1mm}\hspace{-1.5mm}+\hspace{-0.5mm}Rd_{f}N\hspace{-2mm}$\tabularnewline
\hline 
SCMA-RIS-LC & $\hspace{-1.5mm}\hspace{-2mm}\begin{array}{c}
R\left(2d_{f}-1\right)\hspace{-1mm}+\hspace{-1mm}\left(4d_{f}+1\right)\hspace{-1mm}\biggl(UM+\hspace{-1mm}\sum_{\begin{array}{c}
\hspace{-1mm}\hspace{-2mm}r=1\\
\hspace{-1mm}\hspace{-2mm}\grave{U}^{r}\neq0
\end{array}}^{R}\hspace{-1mm}\hspace{-1mm}\hspace{-2mm}M^{\grave{U}^{r}}\hspace{-1mm}\biggr)\hspace{-1mm}+\hspace{-0.5mm}Rd_{f}N+1\hspace{-1mm}\hspace{-2mm}\end{array}$ & $\begin{array}{c}
2Rd_{f}+\left(4d_{f}+2\right)\hspace{-1mm}\biggl(UM+\sum_{\begin{array}{c}
\hspace{-1mm}\hspace{-2mm}r=1\\
\hspace{-1mm}\hspace{-2mm}\grave{U}^{r}\neq0
\end{array}}^{R}\hspace{-1mm}\hspace{-2mm}M^{\grave{U}^{r}}\hspace{-1mm}\biggr)+Rd_{f}N\end{array}$\tabularnewline
\hline 
\end{tabular}
\end{table*}

\textbf{Stage 2}: In this stage, a single user's codeword is detected
at a time using his $d_{v}$ non-zero OREs that carry its codeword.
Also, all other users' codewords are considered to be known from Stage
1 (or from Stage 2 if they have been detected already). The detected
user's codeword, $\hat{\mathbf{c}}_{m}^{u}$, is

\vspace{-3mm}

\begin{equation}
\hat{\mathbf{c}}_{m}^{u}=\hspace{-2mm}\hspace{-2mm}\underset{\begin{array}{c}
j=1,\ldots,M\end{array}}{\text{arg}\,\text{\,min}}\hspace{-2mm}\sum_{r\in\Omega_{u}}\Bigl|y^{r}-\Psi_{3}-\Psi_{4}(j)\Bigr|^{2},\,\,u=1,\,\ldots,\,U,\label{eq: MSUD}
\end{equation}

\vspace{-1.5mm}

\noindent where

\vspace{-3mm}

\begin{equation}
\Psi_{3}=\hspace{-1.5mm}\hspace{-2mm}\sum_{\grave{u}\in\varLambda_{r}\backslash u}\hspace{-2mm}\left(\sum_{n=1}^{N}\alpha_{\grave{u},n}^{r}\beta_{n}^{r}e^{\jmath(\Theta_{\grave{u},n}^{r})}\hspace{-1.5mm}\right)\hspace{-0.5mm}\hat{c}_{\grave{u},m}^{r}\,\Bigl|\,\hat{c}_{\grave{u},m}^{r}\subset\hat{\mathbf{C}}_{\text{LC}},\label{eq: epsi 3}
\end{equation}

\vspace{-3mm}

\begin{equation}
\Psi_{4}(j)=\left(\sum_{n=1}^{N}\alpha_{u,n}^{r}\beta_{n}^{r}e^{\jmath(\Theta_{u,n}^{r})}\right)c_{u,m(j)}^{r}.\label{eq: epsi 4}
\end{equation}

\noindent It is worth noting that $\Psi_{4}$ represents the desired
user's information that needs to be detected. Algorithm \ref{alg: SCMA-RIS-LC}
summarizes the procedure of the SCMA-RIS-LC decoder.

\begin{algorithm}[t]
\begin{itemize}
\item \textbf{Input} channel matrices and codebooks for all users;
\item \textbf{Buffer}\textbf{\small{} }{\small{}$\hat{\mathbf{C}}_{\text{LC}}=\{\cdot\}$}
and \textbf{$\varLambda=\{\cdot\}$};
\end{itemize}
~~~~~1: \textbf{While$\,\,\,r\leq R$, do}

~~~~~2:\textbf{ ~~~Set} $\grave{\varLambda}_{r}\leftarrow\{\varLambda\cap\varLambda_{r}\}$;

~~~~~3: ~~~\textbf{Compute }$\bar{y}^{r}\leftarrow y^{r}-\Psi_{1}$,
using (\ref{eq: epsi 1});

~~~~~4:\textbf{ ~~~Solve }$\{\hat{\mathbf{C}}^{r},\hat{\mathbf{j}}^{r}\}=\underset{\begin{array}{c}
j=1,\ldots,M^{\grave{U}^{r}}\end{array}}{\text{arg}\,\text{\,min}}\Bigl|\bar{y}^{r}-\Psi_{2}(j)\Bigr|^{2}$;

~~~~~5:\textbf{ ~~~Update} $\hat{\mathbf{C}}_{\text{LC}}$
based on $\hat{\mathbf{C}}^{r}$ and $\varLambda$ based on $\hat{\mathbf{j}}^{r}$;

~~~~~6: ~~~\textbf{ if} $\text{card}\{\varLambda\}==U$

~~~~~7:\textbf{ ~~~~~ Go to }Line \#10;

~~~~~8: ~~~\textbf{ end if}

~~~~~9: ~~~\textbf{ Set $r\leftarrow r+1$;}

~~~~10: \textbf{end While}

{\small{}~~~~~11:}\textbf{\textit{\small{} }}\textbf{\textit{For
$k=1:K$, do}}

{\small{}~~~~~12:~~~ }\textbf{\textit{For$\,\,\,u=1:U$, do}}

{\small{}~~~~~13: ~~~~~~}\textbf{ Compute }$\bar{y}^{r}\leftarrow y^{r}-\Psi_{3}$,
using (\ref{eq: epsi 3});

{\small{}~~~~~14:}\textbf{\small{} ~~~~~~~}\textbf{Solve}\textbf{\small{}
}$\hat{\mathbf{c}}_{m}^{u}=\hspace{-5mm}\underset{\begin{array}{c}
j=1,\ldots,M\end{array}}{\text{arg}\,\text{\,min}}\hspace{-2mm}\sum_{r\in\Omega_{u}}\Bigl|\bar{y}^{r}-\Psi_{4}(j)\Bigr|^{2}$;

{\small{}~~~~~15:}\textbf{\small{} ~~~~~~}\textbf{~Update}
$\hat{\mathbf{C}}_{\text{LC}}$ based on $\hat{\mathbf{c}}_{m}^{u}$;

{\small{}~~~~~16:~~~ }\textbf{\textit{end For}}

{\small{}~~~~~17: }\textbf{\textit{end For}}
\begin{itemize}
\item \textbf{Output}{\small{} }$\hat{\mathbf{C}}_{\text{LC}}$.
\end{itemize}
\caption{\label{alg: SCMA-RIS-LC}The SCMA-RIS-LC algorithm pseudo-code.}
\end{algorithm}

\subsection{Complexity Analysis}

In this subsection, the decoding complexity of the proposed SCMA-RIS
decoders is deduced in terms of the real additions (RA) and real multiplications
(RM) that are required to decode the users' codewords. Table \ref{tab:Complexity_SCMA_RIS}
presents the complexity summary for the two proposed decoders of the
SCMA-RIS scheme and the conventional SCMA-MPA. It is worth noting
that the summation term in the RA and RM expressions for the SCMA-RIS-LC
decoder depends on the system parameters; for $U=6$, $R=4$ with
$d_{f}=3$ it becomes $M^{3}+M^{2}+M$. Numerical comparisons will
be given next.

\section{\label{sec:Simulation-Results}Simulation Results}

In this section, Monte-Carlo simulations are used to assess the BER
performance of the proposed SCMA-RIS schemes compared to the conventional
SCMA-MPA scheme. It is assumed that the Rayleigh fading channels of
all users are perfectly known at the receiver. In our setup, we consider
$U=6$, $R=4$ with $d_{f}=3$, $M=4$ and $2$, and $N=20$, $30$
and $40$. Moreover, the assessment of the blind and optimized scenarios
in (\ref{eq: theta}) for the proposed SCMA-RIS schemes is presented.

The number of iterations, $K$, used in MPA for the proposed SCMA-RIS-MPA
and conventional SCMA-MPA is $4$. It is worth noting that the improvement
of the BER performance for the MPA decoder is saturated at $K=4$
(i.e., no improvement occurs when $K>4$) for both SCMA-MPA and SCMA-RIS-MPA.

\begin{figure}[t]
\vspace{-4mm}

\begin{centering}
\includegraphics[scale=0.445]{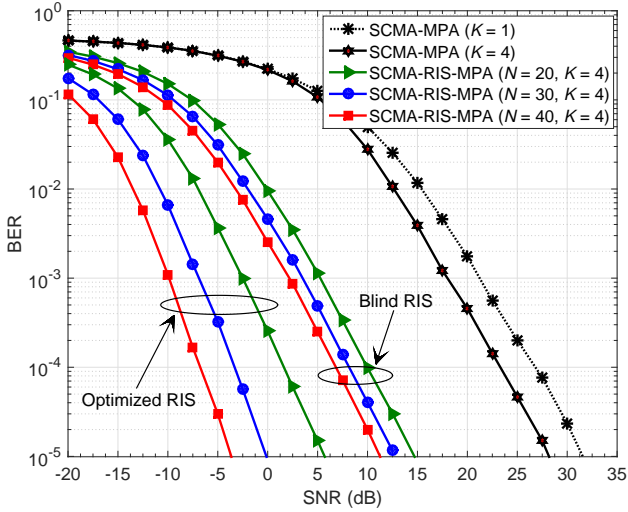}
\par\end{centering}
\vspace{-2.5mm}

\caption{\label{fig: MPA_M2}{\small{}BER performance comparison} of the SCMA-MPA
and SCMA-RIS-MPA schemes for $M=2$.}
\end{figure}

\begin{figure}[t]
\vspace{-4mm}

\begin{centering}
\includegraphics[scale=0.445]{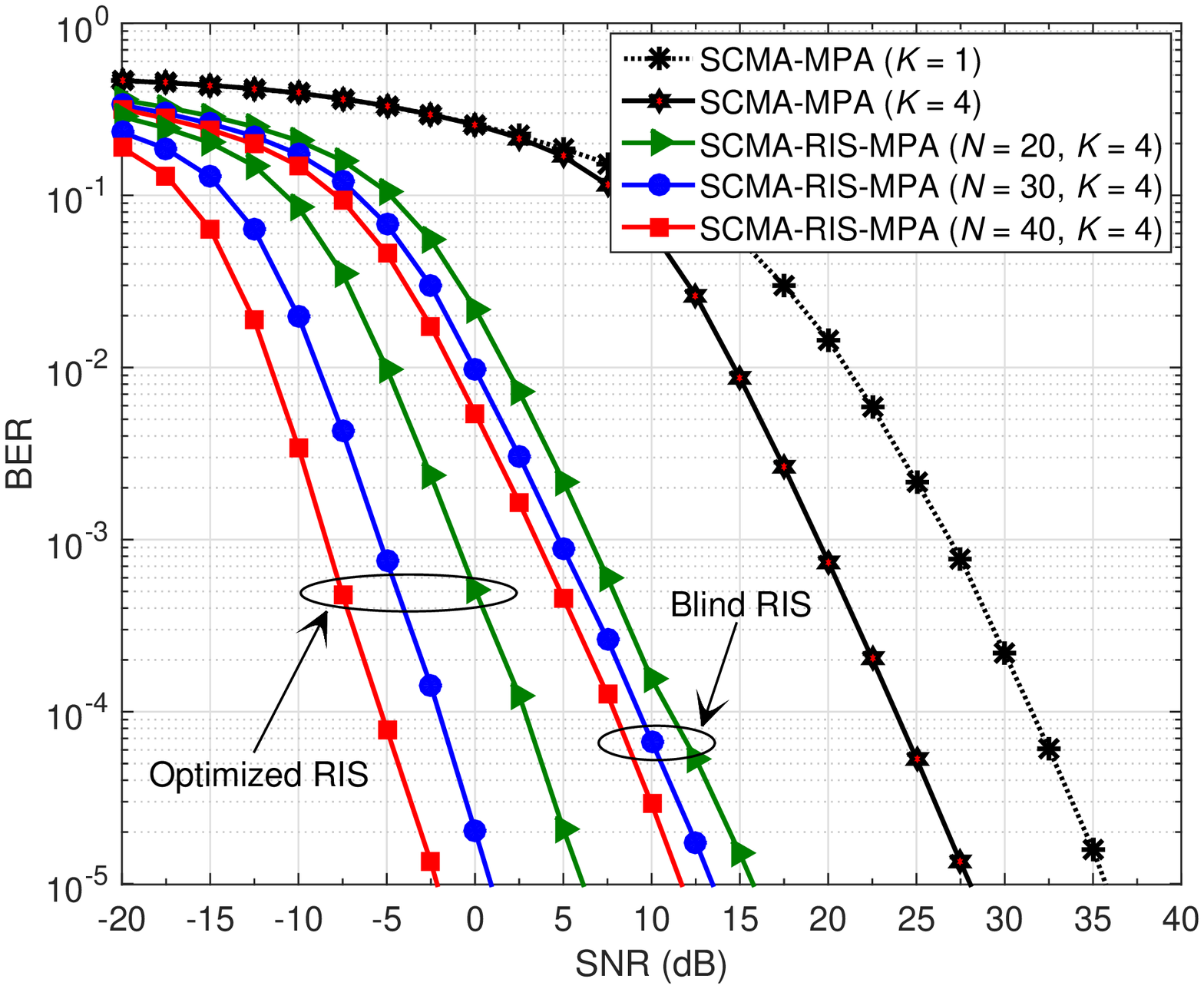}
\par\end{centering}
\vspace{-2.5mm}

\caption{\label{fig: MPA_M4}{\small{}BER performance comparison of }the SCMA-MPA
and SCMA-RIS-MPA schemes for $M=4$.}
\end{figure}

Figs. \ref{fig: MPA_M2} and \ref{fig: MPA_M4} depict the BER performance
comparison between the proposed SCMA-RIS-MPA and conventional SCMA-MPA
for $M=2$ and $4$, respectively. As seen from these figures, the
two SCMA-RIS-MPA scenarios (i.e., blind and optimized) significantly
improve the BER performance compared to the conventional SCMA-MPA.
For instance, the blind SCMA-RIS-MPA scheme for $M=2$ and $4$ provides
around $13$ dB, $15$ dB and $16.5$ dB improvement in the BER performance
for $N=20$, $30$ and $40$ with $K=4$, respectively, compared to
the SCMA-MPA with $K=4$. For the optimized SCMA-RIS-MPA scenario
of $M=2$ and $4$, the improvement in the BER performance increases
to $22$ dB, $27$ dB and $31$ dB for $N=20$, $30$ and $40$ with
$K=4$, respectively, compared to the SCMA-MPA with $K=4$. Thus,
the optimized scenario and an increased number of reflecting elements
provide more BER improvements.

On the other hand, the decoding complexity of the SCMA-RIS-MPA is
slightly higher than for the conventional SCMA-MPA. For example, from
Table \ref{tab:Complexity_SCMA_RIS} for $M=2$ and $K=4$, the increase
in the number of RA and RM of the SCMA-RIS-MPA does not exceed $30.4\%$
and $17.8\%$ than that of the conventional SCMA-MPA, respectively.
Furthermore, when $M=4$, the increase in both RA and RM number for
the SCMA-RIS-MPA does not exceed $3.4\%$.

\begin{figure}[t]
\vspace{-4mm}

\begin{centering}
\includegraphics[scale=0.445]{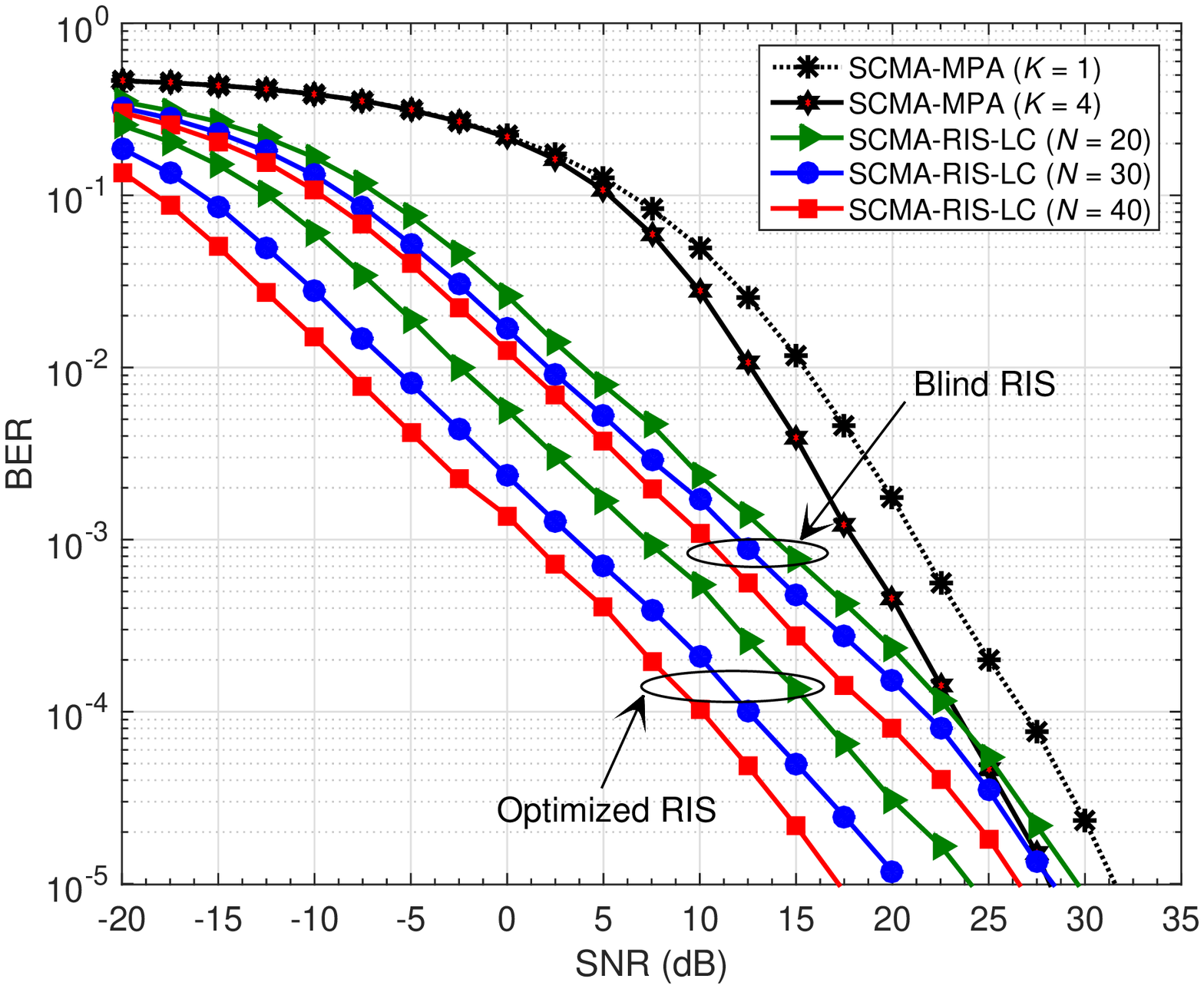}
\par\end{centering}
\vspace{-2.5mm}

\caption{\label{fig: LC_M2}{\small{}BER performance comparison} of the SCMA-MPA
and SCMA-RIS-LC schemes for $M=2$.}
\end{figure}

\begin{figure}[t]
\vspace{-4mm}

\begin{centering}
\includegraphics[scale=0.445]{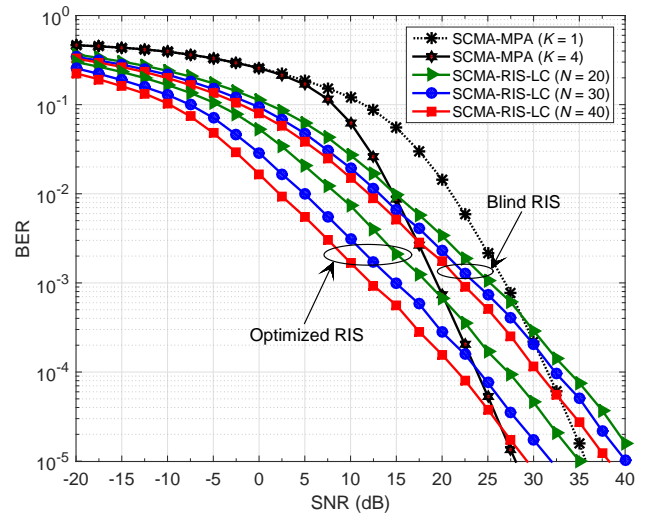}
\par\end{centering}
\vspace{-2.5mm}

\caption{\label{fig: LC_M4}{\small{}BER performance comparison} of the SCMA-MPA
and SCMA-RIS-LC schemes for $M=4$.}
\end{figure}

As seen from Figs. \ref{fig: LC_M2}, \ref{fig: LC_M4} and Table
\ref{tab:Complexity_SCMA_RIS}, the proposed SCMA-RIS-LC decoder significantly
reduces the decoding complexity with a better BER performance than
the SCMA-MPA, especially in low SNR. For instance, the reduction in
the RA for the SCMA-RIS-LC with $M=2$ is $47\%$, $54.6\%$ and $62.2\%$,
while it saves around $68\%$, $72.3\%$ and $76.7\%$ RM for $N=40$,
$30$ and $20$, respectively, compared to SCMA-MPA with $K=4$. For
$M=4$, the saving in the complexity of the proposed SCMA-RIS-LC decoder
enhances compared to the SCMA-MPA with $K=4$; it is around $85.4\%$,
$86.3\%$ and $87.2\%$ in the RA, whereas it reaches $90.4\%$, $91\%$
and $91.5\%$ for RM when $N=40$, $30$ and $20$, respectively.
Thus, as $N$ decreases, the saving in the complexity increases, whereas
the improvement of the BER performance decreases.

Finally, the proposed SCMA-RIS-MPA provides a significant improvement
of the BER performance at the expense of an acceptable increase in
the complexity when compared with the conventional SCMA-MPA. Further,
the proposed SCMA-RIS-LC significantly reduces the complexity with
a considerable improvement in the BER performance when compared with
SCMA-MPA.

\section{\label{sec:Conclusion}Conclusion}

In this letter, we have investigated the SCMA under two RIS scenarios,
referred to as blind and optimized SCMA-RIS. The MPA is adapted to
decode the SCMA-RIS transmitted signals. Furthermore, a low-complexity
decoder is proposed and analyzed for the SCMA-RIS, i.e., SCMA-RIS-LC.
It is shown that the SCMA-RIS-MPA provides up to $31$ dB improvement
in the BER performance with an increase in the decoding complexity
that varies from $30.4\%$ to less than $3.4\%$, compared to the
SCMA-MPA. Furthermore, the proposed SCMA-RIS-LC reduces the decoding
complexity in a range of $47\%$ to $91.5\%$ with a better BER performance,
compared to the SCMA-MPA.


\begin{thebibliography}{10}
\bibitem{Mohammadkarimi_Octavia}M. Mohammadkarimi, M. A. Raza, and
O. A. Dobre, \textquotedblleft Signature-based nonorthogonal massive
multiple access for future wireless networks: Uplink massive connectivity
for machine-type communications,\textquotedblright{} \emph{IEEE Veh.
Technol. Mag.}, vol. 13, no. 4, pp. 40-50, Dec. 2018.

\bibitem{NOMA_Octavia_2016}S. M. R. Islam et al., \textquotedblleft Power-domain
non-orthogonal multiple access (NOMA) in 5G systems: Potentials and
challenges,\textquotedblright{} \textit{IEEE Commun. Surv. Tuts.},
vol. 19, no. 2, pp. 721-742, Oct. 2016.

\bibitem{NOMA_Survey_2017}Z. Ding et al., \textquotedblleft A survey
on non-orthogonal multiple access for 5G networks: Research challenges
and future trends,\textquotedblright{} \textit{IEEE J. Sel. Areas
Commun.}, vol. 35, no. 10, pp. 2181-2195, Oct. 2017.

\bibitem{Low-cost}I. Al-Nahhal, O. A. Dobre, E. Basar, and S. Ikki,
\textquotedblleft Low-cost uplink sparse code multiple access for
spatial modulation,\textquotedblright{} \textit{IEEE Trans. Veh. Technol.},
vol. 68, no. 9, pp. 9313-9317, Jul. 2019.

\bibitem{Nikopour_SCMA_2013}H. Nikopour and H. Baligh, \textquotedblleft Sparse
code multiple access,\textquotedblright{} in \textit{Proc. IEEE Int.
Symposium on Personal Indoor and Mobile Radio Commun. (PIMRC)}, Sep.
2013, pp. 332-336.

\bibitem{codebook_design_2014}M. Taherzadeh et al., \textquotedblleft SCMA
codebook design,\textquotedblright{} in \textit{Proc. IEEE Veh. Technol.
Conf. (VTC Fall)}, Sep. 2014, pp. 1\textendash 5.

\bibitem{MPA_2015}H. Mu, Z. Ma, M. Alhaji, P. Fan, and D. Chen, \textquotedblleft A
fixed low complexity message pass algorithm detector for up-link SCMA
system,\textquotedblright{} \textit{IEEE Wireless Commun}.\textit{
Lett}.\textit{,} vol. 4, no. 6, pp. 585-588, Dec. 2015.

\bibitem{RIS_Ertugrul 2019}E. Basar, M. D. Renzo, J. de Rosny, M.
Debbah, M.-S. Alouini, and R. Zhang, ``Wireless communications through
reconfigurable intelligent surfaces,'' \textit{IEEE Access}, vol.
7, pp. 116753-116773, Sep. 2019.

\bibitem{RIS_2020}M. A. ElMossallamy et al., ``Reconfigurable intelligent
surfaces for wireless \hspace*{-0.5mm}communications: \hspace*{-0.5mm}Principles,
challenges, and opportunities,'' \textit{IEEE Trans. Cogn. Commun.
and Netw}., vol. 6, pp. 990-1002, Sep. 2020.

\bibitem{RIS_3}E. Basar, ``Transmission through large intelligent
surfaces: A new frontier in wireless communications,'' in \textit{Proc.
European Conf. Netw. Commun. (EuCNC)}, Aug. 2019, pp. 112-117.

\bibitem{NOMA_RIS_1}T. Hou, Y. Liu, Z. Song, X. Sun, Y. Chen and
L. Hanzo, ``Reconfigurable intelligent surface aided NOMA networks,''
\textit{IEEE J. Sel. Areas Commun.}, Early Access, 2020, doi: 10.1109/JSAC.2020.3007039.

\bibitem{NOMA_RIS_2}X. Liu et al., ``RIS enhanced massive non-orthogonal
multiple access networks: Deployment and passive beamforming design,''
\textit{IEEE J. Sel. Areas Commun.}, Early Access, 2020, doi: 10.1109/JSAC.2020.3018823.

\bibitem{NOMA_RIS_3}M. Elhattab et al., ``Reconfigurable intelligent
surface assisted coordinated multipoint in downlink NOMA networks,''
\textit{IEEE Commun}.\textit{ Lett}., Early Access, 2020, doi: 10.1109/LCOMM.2020.3029717.
\end{thebibliography}
\end{document}